# SOLAR NEUTRINOS: SENSITIVITY TO

# PRE-MAIN SEQUENCE EVOLUTION AND TO

# THE DEPTH OF THE CONVECTIVE ZONE


John N. Bahcall[1]

*School of Natural Sciences, Institute for Advanced Study, Princeton, New*

*Jersey 08540, USA*

*and*

*Ami Glasner[2]*

*Racah Institute of Physics, Hebrew University, Jerusalem, Israel*

*and*

*Department of Astronomy and Astrophysics,*

*University of Chicago, Chicago*

*IL. 60637, USA*



[1]  *e–mail address: jnb@sns.ias.edu*

[2]  *e–mail address: glasner@vms.huji.ac.il*




## ABSTRACT


Predicted rates for solar neutrino experiments that are obtained with a modified stellar evolution code originally developed to study the advanced stages of stellar evolution are shown to be in agreement with other recently-calculated precise solar models to about 2% (0.2 SNU for the chlorine experiment). Different scenarios for pre-main sequence evolution are explored and are found to change the predicted rates for solar neutrino experiments by less than or of order 1%. The influence of the depth of the solar convective zone on the predicted solar neutrino fluxes is established by direct calculation. It is shown that a change in the calculated depth of the convective zone that is five times larger than the quoted helioseismological measurement uncertainty determined by Christensen-Dalsgaard, Gough, & Thompson causes a change in the predicted $^8$B neutrino flux of less than 7% and a change in the $^7$Be neutrino flux of less than 4%. Additionally, it is shown that the radiative opacities near the depth of the convective zone cannot differ from the standard OPAL values by more than about 12% without causing the calculated and measured depths of the convective zone to differ by more than five times the helioseismological measurement uncertainty.


Subject headings : *– stellar evolution – stellar neutrinos – stars: The Sun – – stars: sun*



# 1. INTRODUCTION

Solar neutrino experiments provide unique information about the interior of the sun and about the nature of neutrinos. In order to make the inferences from these experiments more precise and more informed, it is important to explore a variety of possible uncertainties in the solar models whose predictions are compared with experiments. It is also important to compare the results of precise solar model calculations obtained with different stellar evolution codes in order to isolate, to understand, and to eliminate any possible dependences upon the calculational procedures. This later goal has largely been accomplished in recent years since precise solar models, which yield consistent results, have been calculated by a number of different groups using very different computer codes (see, e.g., Bahcall & Ulrich 1988; Lebreton & Dappen 1988; Sackmann, Boothroyd, & Fowler 1990; Proffitt & Michaud 1991; Guzik & Cox 1991; Bahcall & Pinsonneault 1992, hereafter denoted by BP; Ahrens, Stix, & Thorn 1992; Christensen-Dalsgaard 1992; Guenther, Demarque, Kim, & Pinsonneault 1992; Berthomieu, Provost, Morel, & Lebreton 1993; Turck-Chièze & Lopes 1993; Castellani, Degl'Innocenti, & Fiorentini 1993).

In this paper, we explore two aspects of solar structure and evolution that have not previously been investigated in detail in connection with solar neutrino calculations. We construct solar models, and calculate solar neutrino fluxes, with different assumed pre-main sequence evolutionary histories since this phase of solar evolution is not strongly constrained by direct observations. We also derive the numerical connection between the measured depth of the convective zone and the calculated neutrino fluxes if one is allowed to suppose that the



radiative opacity in the vicinity of the base of the convective zone is significantly in error.

This paper is organized as follows: §2 describes briefly the $ASTRA$ code used to calculate the solar models; §3 focuses on the improvements made in the code in order to carry out the present study; §4 presents the results of our calculations; and §5 summarizes the main conclusions.

## 2. THE EVOLUTIONARY CODE

In this section, we describe some features of the ASTRA stellar evolution code (Rakavy, Shaviv, & Zinamon 1967) that are important for the calculations that are presented in the present paper. This code was originally developed in order to carry out calculations for the advanced stages of stellar evolution of moderate mass stars, but has been adapted here to permit precise calculations of solar evolution. We begin this section by discussing the form of the outer boundary condition, then outline how the code treats convection, and finally summarize the implementation of the equation of state.

The $ASTRA$ code uses as an outer boundary condition the relation between the luminosity of the star and the opacity, pressure and temperature at the outermost zone that is given by Schwarzschild (1958), eq. (11.5). This relation, based on the integration of the hydrostatic equation assuming radiative energy transfer, holds only for radiative envelopes. It also applies, according to figure (11.1) in Schwarzschild (1958) to the radiative regions of the envelope below the convective zone. Hence, in order to describe the convective envelope of the sun, we added an option that uses the same relation at the base of the convective



envelope. We defined a parameter called $T_{convec}$ (replacing the mixing length parameter in codes that use mixing length theory, MLT, for convection). The assumption is that at $T_{convec}$ the envelope is still radiative and that for temperatures below $T_{convec}$ it is fully convective, and therefore also isentropic. In that way we are able to use Schwarzschild's relation and the isentropic relations to find the luminosity as function of the physical parameters at the outermost zone and of the parameter $T_{convec}$. We verified that the solutions are self-consistent, i.e. a model calculated with a certain value of the parameter $T_{convec}$, developed a convective envelope with a temperature at its base of approximately $T_{convec}$. For our best standard solar models, $T_{convec} = 2.23$ and the actual temperature at the base of the convective zone was $T_{base-con} = 2.13 \times 10^6 \mathrm{K}$.

To what extent is the convective zone calculated by our code equivalent to the convective zone computed using mixing length theory? All codes agree in showing that the solar convective zone is isentropic to an excellent approximation. Moreover, all codes give essentially the same results for the inner radiative regions (see BP). The location of the base of the convective zone uniquely defines the entropy of the convective region. The requirement of hydrostatic equilibrium combined with the known entropy determines the structure of the envelope (independent of details of the transport mechanism). Thus, although we do not calculate in detail the structure of the envelope, the quantities we do calculate must agree with the corresponding results obtained using MLT.

The *ASTRA* code (Rakavy, Shaviv, & Zinamon 1967) integrates the time-dependent equations of heat conduction and the rate of composition change while preserving hydrostatic



equilibrium at each step. The independent variables are the entropy and the density at each zone. The code separates the changes in the entropy and the composition from the condition of hydrostatic equilibrium. At each time step, the hydrostatic equations are solved for a given fixed entropy profile by a quasi-dynamic method.

Convection is treated according to the following three-part procedure. First, the entropy, $S$, is changed at each grid point by an amount proportional to an infinitesimal time step, $\delta t$, using an explicit "radiative" formula:

$$\text{new } S_i = \text{old } S_i + [\frac{(\text{Flux}_{i-1} - \text{Flux}_i)}{\Delta M} + q_i] \times \frac{\delta t}{T_i}, \tag{1}$$

where Flux, $\Delta M$, $T$, and $q$ are, respectively, the integrated radiative flux (erg/sec), the zone mass, temperature and nuclear energy generation rate. Second, the borders of the new convective zones are determined according to the Schwarzschild criteria in the following way: regions with homogeneous composition are convectively unstable if the entropy gradient is negative. In regions where composition changes from one zone to the other, matter from zone $i$ is tested for convective stability with zone $i + 1$ by assuming that matter from zone $i$ that carries with it the specific entropy $S_i$ and abundances $X_i$, floats and equalizes pressure with zone $i + 1$ so that the pressure is $p_{i+1}$. Under those conditions the temperature of this "blob" is determined by the equation of state to be $T_{blob} = T(p_{i+1}, S_i, X_i)$. The zone is unstable to convection if $T_{blob} > T_i$. This test is performed twice at each boundary: once for a floating "blob" (the pair $i, i+1$) and once for a sinking "blob" (the pair $i+1, i$). Third, the composition and entropy are averaged within a convective zone. If all the intervals which



form a convective zone have identical composition, the entropy of the convective zone is defined as the mass average of all intervals within the zone. If the original composition is not homogeneous, the compositions are averaged. The internal energy of the entire convective zone before mixing is calculated. Then iterations are performed until we find the proper specific entropy $< S_c >$ of the mixed convective zone, so that the energy of the entire zone is conserved, i.e. equals to $< U_c >$.

$$< U_c > = \Sigma \Delta M_i \times u_i \qquad (2)$$

where $u_i$ is the specific internal energy of the i-th zone and the summation is over all mass intervals in the convective zone. Comparing this treatment of convection with the commonly used prescription of the mixing length theory, we can state that the criteria for determination of convective stability are essentially the same. As to the treatment of the convective zone itself, the $ASTRA$ treatment assumes immediate equalization of entropy, instead of (as in conventional mixing length theory) a convective flux that tends to make the convective region isentropic. The assumption of equal entropy is appropriate since the convective overturn time of a "blob" is much shorter than the characteristic evolutionary time. The convective regions in Bahcall & Ulrich (1988) and in BP are isentropic, just as in the $ASTRA$ code. The ideal gas entropy $S = 1./\mu \times ln(T^{1.5}/\rho)$, where $1./\mu = (5 \times x + 3)/4$, is 38.85 for Bahcall & Ulrich (1988), 39.32 for BP, and 38.72 for our Model 2.

For the equation of state, $ASTRA$ uses a subroutine from the original (Rakavy, Shaviv, & Zinamon 1967) code that includes electron degeneracy and relativistic effects. The only



change made in the equation of state for the purposes of this study was the addition of a term that corrects the pressure for Debye-Huckel interaction, which was included according to Equation (14) of BP.

## 3. IMPROVED ASPECTS OF THE CODE

In this section, we summarize improvements that have been made in the modified AS-TRA code regarding the input radiative opacities, the nuclear-burning subroutine, and the elemental abundances. These improvements allow us to calculate accurate solar neutrino fluxes and to compare our model results with other recent precision solar model calculations.

The radiative opacities used in this application of the modified ASTRA code are the OPAL values (Iglesias & Rogers 1991a,b,c; Rogers & Iglesias 1992; see Table VI of BP) for the Anders & Grevesse (1989) mixture of heavy elements, assuming the meteoretic iron abundance. The opacities and element abundances used here represent significant improvements over previously used values (see discussion in BP).

Nuclear burning and neutrino production are calculated using the Export version of the energy-generation routine energy.for (described by Bahcall & Pinsonneault 1992); this sub-routine was implemented within the $ASTRA$ code. The subroutine represents the nuclear reaction rates in the standard form used by the experimentalists who determine rate parameters and provides as default values for nuclear rates the most recently-established cross section factors.

The isotopic abundances are solved for implicitly at each time step using the rates



given by the energy routine. The relevant elements in our network of nuclear reactions are: $^1$H, $^3$He, $^4$He, $^{12}$C, $^{13}$C, $^{14}$N, $^{16}$O, and $^{18}$O. We used the nuclear cross-section parameters given in Table 1, column 7, of Bahcall & Pinsonneault (1992). We included weak and intermediate nuclear screening according to the prescription of BP.

All other input parameters, such as solar luminosity ($3.86 \times 10^{+38}$ erg/sec), age ($4.6 \times 10^{+9}$ years), and radius ($6.97 \times 10^{+10}$ cm), are the same as in BP.

In order to converge the solar models, we iterated an evolutionary series of models up to the present age of the sun by changing the metallicity mass fraction, Z, while keeping the ratio $Z/X$ constant at the observed (Anders & Grevesse 1989) value of $Z/X = 0.02671$. All models had a total of 220 zones with a central zone of $10^{-5} M_\odot$ and an outermost zone of $5 \times 10^{-4} M_\odot$.

## 4. THE SOLAR MODELS

This section focuses on the effect of pre-main sequence evolution on the calculated solar neutrino fluxes (§4.1), the relation between the depth of the convective zone and the calculated neutrino fluxes (§4.2), and the comparison between the standard model results obtained here and in previous accurate calculations of solar neutrino fluxes (§4.3).

### 4.1 Pre-Main Sequence Evolution

The sensitivity of models of the current sun to pre–main sequence evolution is expected to be negligible (Iben 1965; Stahler 1994). In order to verify that this is the case also for the



computed solar neutrino fluxes (which are known to be sensitive some input parameters), we compare in this sub-section two models with very different pre-main sequence histories. In the first model, elemental abundances are kept constant during pre-main sequence evolution while in the second model composition changes are computed using the Export version of the nuclear burning subroutine of BP.

In order to construct the simplest model, Model 1 in Table 1, we started with an initial isentropic profile of a solar model on the Hayashi track ( $T_c = 2.0 \times 10^{+6}$ K, $\rho_c = 0.15$ g cm$^{-3}$ and $R_{\rm out} = 3.85 R_\odot$) The evolution equations were first solved without allowing nuclear burning to change the composition, using nuclear rates only as an energy source. The star loses entropy, contracts, and approaches thermal equilibrium. In this way, the profile we obtain is identical to a profile one would obtain for fixed initial abundances by solving the hydrostatic equation and demanding thermal equilibrium, i.e: the total luminosity is equal to the integrated nuclear energy generation rate. The epoch at which thermal equilibrium is achieved is defined as the zero age main sequence (ZAMS). Following the ZAMS, we include changes in the nuclear abundances. We evolve the models with small time steps making only two or three iterations each step. At each step iterations convergence if the entropy changes, in units of the gas constant ( $k \times N_0$), by less than 0.0001 from one iteration to the next. In general 250 time steps were required to advance from the ZAMS to the present age. A model was also calculated with only 70 time steps. It required many more iterations per time step.

The principal characteristics of Model 1 are given in Table 1. Table 1 presents the present-day central hydrogen and helium abundances, $X_c, Y_c$, the inferred primordial helium



abundance, $Y_{pri}$, the heavy-element abundance (assumed uniform), $Z$, the calculated central temperature and density, $T_c$ and $\rho_c$, and the calculated solar radius and luminosity, $R_{outer}$ and $L$. The calculated neutrino fluxes and the predicted capture rates for the chlorine and gallium solar neutrino experiments (in solar neutrino units, SNU) are given in Table 2. The neutrino fluxes are given in units of $cm^{-2}s^{-1}$ at the earth's surface with the common power of ten removed and indicated in the columns referring to each neutrino source. For comparison, we list the characteristics of the best solar model without diffusion of Bahcall and Pinsonneault in the row labeled Model 5 of Table 1–Table 3. The agreement is excellent and will be discussed in more detail in § 4.3.

In order to investigate the sensitivity of our results to the details of the pre-main sequence evolution, we calculated a second model, Model 2 in Table 1. This model was constructed using the same assumed initial entropy profile as for Model 1, but for Model 2 nuclear burning was allowed to change the abundances on the way to thermal equilibrium. Again, the star contracts, loses entropy, and approaches thermal equilibrium. When thermal equilibrium is achieved, the abundances at the center of the star differ from their original values. For example, the hydrogen abundance decreased from 0.7081 to 0.7074 and the $^{12}C$ abundance decreased from 0.00305 to 0.00166.

The present-day characteristics of the sun in Model 2 are almost identical to the characteristics of the simpler model, Model 1. For example, the difference in the $^8B$ neutrino flux calculated for Model 1 and for Model 2 is only about 1%.

We also calculated two other solar models one in which the initial central temperature



was twice as high as in Model 1, i.e., $T_c = 4.0 \times 10^{+6}$ K, and the other in which the density

of each zone in the initial model was decreased (with respect to Model 1) by a multiplicative

factor ranging linearly in mass with a value of 0.8 in the center and 0.1 at the surface, so that

the initial model was not isentropic. Both models were evolved in the same way as Model 1.

As expected, the characteristics of the converged models were essentially indistinguishable

from the characteristics of Model 1. For example, the calculated neutrino fluxes were the

same–to the accuracy shown in Table 2–as the neutrino fluxes of Model 1, except for the $^8$B

neutrino flux which differed by at most 0.2% from the value given in Table 2.

## 4.2 *The Depth of the Convective Zone*

We explore in this section the dependence of the computed neutrino fluxes on the calcu-

lated depth of the convective zone.

By choosing different values for the convective parameter $T_{\mathrm{convec}}$ (defined in § 2) and by

iterating the assumed initial metallicity, we tried to obtain different depths for the convective

zone. We were unable to find satisfactory solutions in which the radius at the base of the

convective zone was more than 2% different from the standard value, as long as the usual

input parameters for the solar model were kept constant. The calculated solar radius for

models that converged to the measured solar luminosity at the present age differed from the

observed solar radius by a few percent in all the trials we made for which $T_{\mathrm{convec}}$ differed by a

few percent from the standard value. These calculations verify that the computed depth of

the convective zone is essentially uniquely specified by other input data to the solar model.

Models based on the MLT would come to the same conclusions. The reason is that



in both cases we have a one-parameter model that responds in the same way when the parameter is varied. The stellar radius decreases with increasing $T_{convec}$ (mixing length) and increases when those parameters are decreased.

In order to evaluate the relationship between the predicted neutrino fluxes and the depth of the solar convective zone, we need to construct solar models with different depths of the convective zones that also correctly reproduce all other (non-helioseismological) observable parameters, including the solar radius. In order to calculate a series of such models, we were forced to change the most uncertain physical quantity that is directly related to the convective stability, i.e., the radiative opacity at temperatures close to the temperature at the base of the convective zone.

The opacity variations were performed in the following way. For temperatures higher than $8.0 \times 10^{+6}$ K, the radiative opacity was not altered. For temperatures lower than $3.0 \times 10^{+6}$ K, the radiative opacity was multiplied by a factor of: $\eta = (1 + f)$, where $f$ is a numerical parameter. For temperatures between $3.0 \times 10^{+6}$ K and $8.0 \times 10^{+6}$ K, a linear interpolation was made in order to produce a smoothly-varying but altered opacity. In order to change the calculated radius at the base of the convective zone by a fractional amount of $0.03$, we found that we were forced to change the radiative opacity by an order-of-magnitude larger fraction, i.e., $f = \pm 0.25$.

Model 3 and Model 4 were constructed in the same way as Model 1, except that for Model 3 the parameter $f = +0.25$ (which produces a deeper convection zone) and for Model 4 the parameter $f = -0.25$ (which produces a shallower convection zone). Note



that the assumed opacity change is relatively large. Iglesias & Rogers (1991a) found that the much-improved OPAL opacities differ by only about 2.5% in the solar interior from the earlier Los Alamos values. In Figure 1, the perturbed radiative opacities (with $f = \pm 0.25$) in Model 3 and Model 4 are compared with the OPAL opacity profile in the standard solar model, Model 2. The opacity changes assumed here are substantial for solar radii larger than about $0.35 \ R_{\odot}$.

The temperature profiles of Models 2, 3, and 4 are compared in Figure 2. The differences in temperatures in the outer regions of the sun are those expected as direct results of the opacity changes. The temperature differences initially decrease as one moves further into the sun. However, close to the center, in the nuclear burning zone, the differences increase again and approach the differences of the central values that are given in Table 1. This behavior can be understood in the following way: in order to deepen the convective zone but still have a solar model that has the same outer parameters as the observed sun, we had to increase the radiative opacities and at the same time decrease the metallicity $Z$ relative to the standard Model 1. Since by construction all of the solar models have the same (observed) primordial value of $Z/X$ at the surface, the initial hydrogen abundance was also reduced in the model with the artificially deepened convective envelope, Model 3. The effect of this slight decrease in hydrogen abundance is a slightly higher nuclear burning temperature, which further reduces the hydrogen abundance. The higher burning temperature increases the $^{7}$Be and $^{8}$B neutrino fluxes, since these fluxes increase as the central temperature increases (see Bahcall 1989, Chapter 6). For the same reasons, the opposite trends appear for a shallower



convective zone. The pp, pep, and hep neutrino fluxes are less sensitive to the central temperature profile and their calculated fluxes are almost unaffected by the assumed changes in the radiative opacity.

Table 3 presents the calculated results relating to the convective zone. The first column gives, as before, the model number, the second column lists the value of the opacity fudge-factor, $f$, the third and fourth columns present the calculated radius at the base of the convective zone, $R_{\text{base conv}}$, and the calculated temperature at the base of the convective zone, $T_{\text{base conv}}$.

For future reference, we record in Table 4 the calculated logarithmic derivatives of the neutrino fluxes with respect to the calculated depth of the convective zone.

The value of the radius at the base of the convective zone has been determined by Christensen-Dalsgaard, Gough, & Thompson (1991) from helioseismological measurements to be:

$$R_{\text{base conv}} = (0.713 \pm 0.003) R_{\odot}. \tag{3}$$

The calculated depth of the convective zone in Models 2 and 3, $R_{\text{base conv}} = 0.725\ R_{\odot}$ and $R_{\text{base conv}} = 0.726\ R_{\odot}$, is in excellent agreement with the value inferred from helioseismology if one uses the result of Bahcall and Pinsonneault (1992) that helium diffusion reduces the calculated radius at the base of the convective zone by approximately 0.014 $R_{\odot}$, i.e.,

$$R_{\text{base conv}}(\text{with He diffusion}) = R_{\text{base conv}}(\text{without He diffusion}) - 0.014\ R_{\odot}\ .$$



Making the correction for helium diffusion, the estimated depth of the convective zone in models like Model 2 and Model 3 is, respectively, $0.711\ R_\odot$ and $0.712\ R_\odot$.

Conservatively, if we require that the calculated depth differ from the measured depth by less than five times the uncertainty estimated by Christensen-Dalsgaard et al., then the allowed uncertainty $\delta R_{\text{base conv}}$ in the depth of the convective zone is

$$\delta R_{\text{base conv}} = 0.015 R_\odot. \tag{4}$$

Comparing the results for Model 3 and Model 4 in Table 3 with the helioseismological measurement, we see that the discrepancy between measured and calculated depths is, after taking account of the correction caused by diffusion,

$$|R_{\text{base conv; meas}} - R_{\text{base conv; calc}}| \leq 0.03\,(f/0.25)\,R_\odot. \tag{5}$$

Table 2 presents the calculated neutrino fluxes and the predicted event rates for the chlorine and for the gallium experiments. Using the results shown in Table 2, we see that the upper limit change in the $^8$B neutrino flux, which causes a change five times larger than the measuring uncertainty of the depth of the convective zone, is

$$\frac{\Delta\phi(^8B)}{\phi(^8B)} \leq 0.07. \tag{6}$$

For the $pp$ neutrino flux, the corresponding upper limit is

$$\frac{\Delta\phi(pp)}{\phi(pp)} \leq 0.004. \tag{7}$$



This reasoning can be reversed and used to place a limit on expected changes in the radiative opacity. Our results imply that the radiative opacity near the base of the convective zone is relatively accurate. If this were not the case, the agreement would be less good between the calculated and the measured depths of the convective zone. If the radiative opacity is changed by more than about 12% from the OPAL value, then the calculated depth of the convective zone would disagree with the measured value by about five times the quoted uncertainty determined by Christensen-Dalsgaard, Gould, & Thomson (1991).

### 4.3 Comparison with Other Standard Models

Model 2 represents our best standard model. This model can be compared with other standard solar models that use the same input parameters by making use of the discussion in §IV of Bahcall & Pinsonneault (1992). These later authors compared their results with solar models by Bahcall & Ulrich (1988), by Sienkiewicz et al. (1990), by Turck-Chièze et al. (1988), and by Sackmann, Boothroyd, & Fowler (1990), all of whom used different stellar evolution codes. When corrected for the differences in input data used by the different groups, all of the codes were found to be in agreement to with ±0.1 SNU. Similar results have been obtained more recently by Berthomieu et al. (1993) and by Castellani et al. (1993).

The comparison between different models is, by design, very direct in the present case: we used the same input parameters as Bahcall & Pinsonneault (1992), precisely in order to simplify the discussion. The results to be compared are given in row 2 (Model 2) and row 5 (Model 5) of Table 1–Table 2. The central temperatures of our best standard model and the central temperature of the BP best model agree to within 0.1%. The predicted event



rates for the chlorine solar neutrino experiment agree to within 0.2 SNU (about 2%) and the predicted event rates for the gallium solar neutrino experiment agree to within 1 SNU (i.e., more closely than 1%).

We conclude that the modified ASTRA code produces standard solar models that are consistent with those obtained with other state-of-the-art codes that do not include element diffusion.

## 5. SUMMARY AND CONCLUSIONS

We have modified the ASTRA code by including an improved nuclear energy generation routine (the Export subroutine of Bahcall & Pinsonneault 1992), more accurate (OPAL) radiative opacities, an equation of state that takes account of the Debye-Huckel correction, and a phenomenological description of the transition between radiative and convective equilibrium. We have verified that this improved code yields results that are in agreement with other recent precise solar models to an accuracy of better than 2% in the predicted rates of solar neutrino experiments (e.g., to an accuracy of 0.2 SNU for the $^{37}$Cl experiment).

We have explored the dependence of the predicted neutrino event rates on the assumed pre-main sequence evolution. We find that different assumptions about the initial entropy or the amount of nuclear burning during the pre-main sequence phase change the calculated rates for current-day solar neutrino experiments by less than or of order 1% (i.e., by less than 0.1 SNU in the $^{37}$Cl experiment).



We attempted to construct standard solar models with depths of the convective zone that are significantly different from the value obtained with the best input parameters. We found that the calculated depth of the convective zone is essentially uniquely determined by the values of the other measured solar parameters and standard input data. We found that significant changes in the calculated depth of the convective zone are most easily accomplished by changing the input radiative opacity near the base of the convective zone.

If the calculated depth of the convective zone is allowed to differ from the standard calculated value by five times the helioseismological measurement uncertainty (Christensen-Dalsgaard, Gough, & Thompson 1991), then the calculated $^8$B neutrino flux can differ by at most 7 percent from the standard calculated flux and the calculated $^7$Be neutrino flux can differ by at most 4 percent from the standard calculated value. Thus if we artificially change the input radiative opacity at the base of the convective zone by approximately 12 percent (a large difference), we can change the predicted event rate in the chlorine solar neutrino experiment by approximately the same percentage.

We conclude that uncertainties produced in the calculated solar neutrino fluxes by either the uncertain nature of pre-main sequence evolution or by the depth of the convective zone are significantly smaller than previously-estimated uncertainties in the neutrino flux calculations (see Bahcall 1989, Chapter 7).

Our results also place an upper limit on the changes in the radiative opacity near the base of the convective zone that may be expected in the future. Changes larger than 12% in the opacity near the base of the convective zone would lead to a serious disagreement,



a disagreement five times larger than the estimated measuring uncertainty, between the calculated depth and the depth determined from helioseismological measurements.

We are grateful to Z. Barkat and to B. Paczyński for valuable discussions. This work was supported in part by NSF grant PHY-91-06210 with the Institute for Advanced Study and by NSF grant AST-92-17969 at the University of Chicago.



TABLE 1

The Basic Parameters

($T$ $\rho$ in c.g.s )

| Model | $X_c$ | $Y_c$ | $Y_{pri}$ | Z | $T_c$ ($10^7$ K) | $\rho_c$ (gm cm$^{-3}$) | $R_{out}/R_\odot$ | $L/L_\odot$ |
|---|---|---|---|---|---|---|---|---|
| 1 | 0.357 | 0.624 | 0.2734 | 0.01888 | 1.555 | 150.7 | 0.999 | 1.00004 |
| [a]2 | 0.352 | 0.628 | 0.2729 | 0.01889 | 1.556 | 151.8 | 1.000 | 1.00007 |
| 3 | 0.345 | 0.636 | 0.2824 | 0.01865 | 1.564 | 154.5 | 1.000 | 0.99994 |
| 4 | 0.371 | 0.609 | 0.2622 | 0.01917 | 1.544 | 145.5 | 1.002 | 1.00006 |
| [b]5 | 0.354 | 0.627 | 0.2716 | 0.01895 | 1.557 | 151.3 | 1.000 | 1.0000 |

[a] Composition changes due to nuclear reactions before thermal equilibrium. See text.

[b] The standard model without diffusion of (Bahcall & Pinsonneault 1992).



TABLE 2

The Neutrino Fluxes

| Model | pp | pep | $^7Be$ | $^8B$ | $^{13}N$ | $^{15}O$ | $^{17}F$ | Cl | Ga |
|---|---|---|---|---|---|---|---|---|---|
| | (E10) | (E8) | (E9) | (E6) | (E8) | (E8) | (E6) | Rate | Rate |
| | $cm^{-2}s^{-1}$ | $cm^{-2}s^{-1}$ | $cm^{-2}s^{-1}$ | $cm^{-2}s^{-1}$ | $cm^{-2}s^{-1}$ | $cm^{-2}s^{-1}$ | $cm^{-2}s^{-1}$ | (SNU) | (SNU) |
| 1 | 6.02 | 1.43 | 4.57 | 4.90 | 4.25 | 3.58 | 4.48 | 7.0 | 126 |
| 2 | 6.02 | 1.43 | 4.60 | 4.96 | 4.26 | 3.62 | 4.54 | 7.0 | 126 |
| 3 | 5.99 | 1.44 | 4.85 | 5.48 | 4.57 | 3.91 | 4.92 | 7.7 | 130 |
| 4 | 6.06 | 1.42 | 4.24 | 4.27 | 3.90 | 3.22 | 4.00 | 6.2 | 122 |
| 5 | 6.04 | 1.43 | 4.61 | 5.06 | 4.35 | 3.72 | 4.67 | 7.2 | 127 |

TABLE 3

Results for Convective Zone

| Model | $f$ | $R_{\text{base conv}}/R\odot$ | $T_{\text{base conv}}$ |
|---|---|---|---|
| 1 | 0.00 | 0.725 | 2.13E+6 |
| 2 | 0.00 | 0.726 | 2.13E+6 |
| 3 | +0.25 | 0.704 | 2.37E+6 |
| 4 | −0.25 | 0.757 | 1.80E+6 |
| 5 | 0.00 | 0.721 | 2.15E+6 |



TABLE 4

Logarithmic Derivatives of Neutrino Fluxes

With Respect to Depth of the Convective Zone

| Model | pp | pep | hep | $^7Be$ | $^8B$ | $^{13}N$ | $^{15}O$ | $^{17}F$ |
|-------|-------|-------|-------|-------|-------|-------|-------|-------|
| 3 | +0.19 | −0.18 | +0.59 | −2.04 | −3.85 | −2.52 | −3.05 | −3.24 |
| 4 | +0.15 | −0.16 | +0.45 | −1.75 | −3.21 | −1.96 | −2.44 | −2.61 |




## References

Ahrens, B., Stix, M., & Thorn, M. 1992, A&A, 264, 673

Anders, E., & Grevesse, N. 1989, Geochim. Cosmochim. Acta, 53, 197

Bahcall, J. N. 1989, Neutrino Astrophysics (Cambridge: Cambridge Univ. Press)

Bahcall, J. N., & Pinsonneault, M. H. 1992, Rev. Mod. Phys., 64, 885

Bahcall, J. N., & Ulrich, R. K. 1988, Rev. Mod. Phys., 60, 297

Berthomieu, G., Provost, J., Morel, P., & Lebreton 1993, A&A, 268, 775

Castellani, V., Degl'Innocenti, S., & Fiorentini, G. 1993, A&A, 271, 601

Christensen-Dalsgaard, J. 1992, Geophys., Astrophys., Fluid Dynamics, 62, 123

Christensen-Dalsgaard, J., Gough, D. O., & Thompson, M. J. 1991, ApJ, 378, 413

Guenther, D. B., Demarque, P., Kim, Y. C., & Pinsonneault, M. H. 1992, ApJ, 387, 372

Guzik, J. A., & Cox, A. N. 1991, ApJ, 381, L333

Iben, I. 1965, ApJ, 141, 993

Iglesias, C. A., & Rogers, F. J. 1991a, ApJ, 371, 408

———. 1991b, ApJ, 371, L73

———. 1991c, private communication

Lebreton, Y., & Däppen, W. 1988, in Seismology of the Sun and Sun-like Stars, ed. V. Domingo, & E. J. Rolfe (ESP SP-286), 661

Proffitt, C. R., & Michaud 1991, ApJ, 380, 238

Rakavy, G., Shaviv, G., & Zinamon, Z. 1967, ApJ, 150, 131

Rogers, F. J., & Iglesias, C. A. 1992, ApJS, 79, 507





Sackmann, I. J., Boothroyd, A. I., & Fowler, W. A. 1990, ApJ, 360, 727

Schwarzschild, M. 1958, Structure and Evolution of the Stars (Princeton: Princeton Univ.

    Press)

Sienkiewicz, R., Bahcall, J. N., & Paczyński, B. 1990, ApJ, 349, 641

Stahler, S. W. 1994, PASP, 106, 337

Turck-Chièze, S., Cahen, S., Cassé, M., & Doom, C. 1988, ApJ, 335, 415

Turck-Chièze, S., & Lopes, I. 1993, ApJ, 408, 347




Figure Captions:

1. The logarithm of the radiative opacity cm$^2$g$^{-1}$ *vs.* the solar radius, for Model 2, Model 3, and Model 4 (full line, dots, broken line). The opacities for Model 1 and Model 2 are so nearly identical that the differences would not be discernable in the figure.

2. Profiles of the temperature $[K]$ *vs.* the solar radius for Model 2, Model 3, and Model 4 (full line, dots, broken line). The temperature profiles for Model 1 and Model 2 are so nearly identical that the differences would not be discernable in the figure.



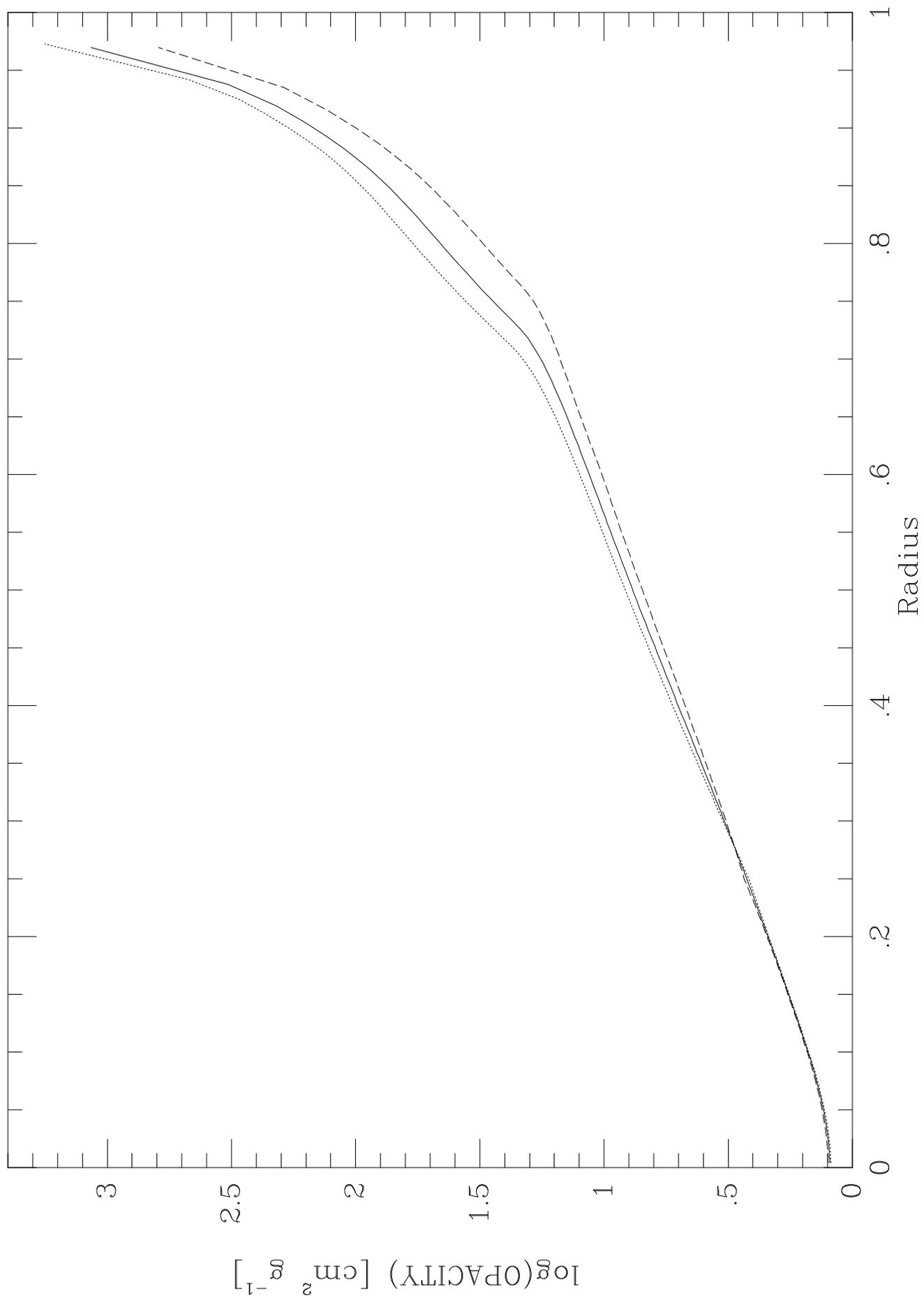

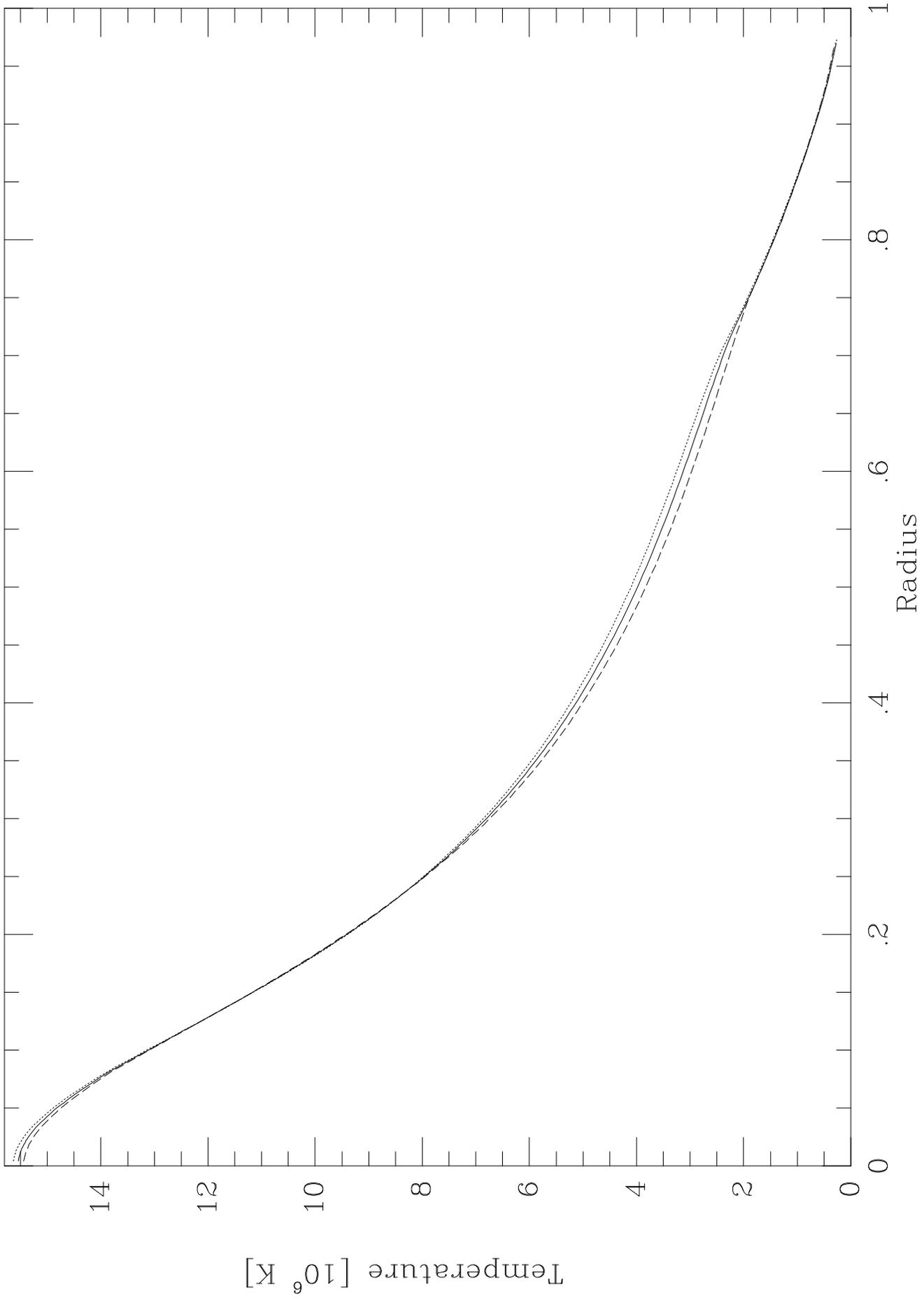